\documentstyle[12pt]{article}
\begin{document}
           \title{\bf Extra force in Kaluza-Klein gravity theory }
\author{ W. B. Belayev \thanks{dscal@ctinet.ru}\\
\normalsize Center for Relativity and Astrophysics,\\ \normalsize
185 Box , 194358, Sanct-Petersburg, Russia }

\maketitle

\begin{abstract}
In induced matter Kaluza-Klein gravity theory the solution of the
dynamics equations for the test particle on null path leads to
additional force in four-dimensional space-time. We find such
force from five-dimensional geodesic line equations and apply this
approach to analysis of the asymmetrically warped space-time.
\end{abstract}

\begin{description}
\item[\it  Keywords:]
induced matter, geodesics
\item[\it  PACS:]
04.50.+h, 04.20.Jb, 04.80.Cc
\end{description}

Five-dimensional space-time theories, including Kaluza-Klein
gravity and brane world, analyze extension of 4D space-time.
Recently studies of the particle movement in 5D have revealed the
consequent departure from 4D geodesic motion by geometric force
using some several approaches for its determination [1-7].  In [1]
one is found by parametrical differentiating of 4D normalization
condition, which is argued by the possibility of the maintenance
relation $f^{i}u_{i}=0$, where $f^{i}$ is component of the extra
force (per unit mass) and $u^{i}$ is 4-velocity. Contrary to this
belief principle of the classic general relativity, proposed that
free particle moves along his geodesic line, is extended on 5D
movement of massive [2-4] and massless [5,6] particles. In induced
matter Kaluza-Klein (IM-KK) gravity theory massive particles in 4D
have not rest mass in 5D and move on null path [1,5,7].

 In this paper
through 5D null geodesic motion analysis in IM-KK gravity theory
it was found out appearance extra forces in 4D, when the scalar
potential depends from coordinates of 4D space-time, and cylinder
conditions failed i.e. metric coefficients depend on the fifth
coordinate. We analyze example of asymmetrically warped
space-time, meaning that the space and time coordinates have
different warp factors.

We will consider typical induced matter scenario, when the 5D
interval is given by
\begin{equation}\label{f1}
dS^{2}  = ds^{2} {\mathrm{-}} \Phi^{2}(x^{m},y)dy^{2} ,
\end{equation}
where $ds$ is 4D line element, $\Phi$ is scalar potential depended
from 4D coordinates $x^{m}$ and extra space-like coordinate $y$.
The 4D line element is written form
\begin{equation}\label{f2}
ds^{2}  = g_{ij}(x^{m},y)dx^{i}dx^{j} ,
\end{equation}
where $g_{ij}$ is metric tensor.

  Since massive in 4D particles move on null path i. e. dS=0, 5D particle dynamics
equations are found for null geodesic line just as in 4D by
extrimizing \cite{a5} function
\begin{equation}\label{f3}
 I=\int\limits_{a}^{b}d\lambda\{g_{ij}\frac{dx^{i}}{d\lambda}
 \frac{dx^{j}}{d\lambda}-\Phi^{2}\frac{dy^{2}}{d\lambda^{2}}\}
 \equiv\int\limits_{a}^{b}hd\lambda  ,
\end{equation}
where $\lambda$ is affine parameter along the path of the particle
terminated at the points $a$, $b$. Null geodesic line equations
are given by
\begin{equation}\label{f4}
\frac{d^{2}X^{A}}{d\lambda^{2}}+\Gamma_{BC}^{A}\frac{dX^{B}}{d\lambda}
 \frac{dX^{C}}{d\lambda}= 0,
\end{equation}
where $X^{A}$ are coordinates of 5D space-time, and
$\Gamma_{BC}^{A}$ are appropriate defined Christoffel symbols. It
should be noticed that if we extremized action with Lagrangian
$L\equiv h^{1/2}$ for particle moving on null path in 5D we would
obtain division by zero, since this movement assigns $h=0$.
Generally choice of parameter $\lambda$ is not arbitrary, and
turned to differentiation with respect to $s$ in Eq. (\ref{f4})we
obtain
\begin{equation}\label{f5}
\frac{d^{2}X^{A}}{ds^{2}}+\Gamma_{BC}^{A}\frac{dX^{B}}{ds}
\frac{dX^{C}}{ds}= -\omega\frac{dX^{A}}{ds},
\end{equation}
where
$\omega=\frac{d^{2}X^{A}}{d\lambda^{2}}/\left(\frac{ds}{d\lambda}\right)^{2}$
and interval $ds$ is assumed to be timelike.

  The first four components of Eq. (\ref{f5}), corresponded to the motion in 4D spacetime, are
transformed to
\begin{equation}\label{f6}
\frac{Du^{i}}{ds}\equiv\frac{d^{2}x^{i}}{ds^{2}}+\Gamma_{jk}^{i}
\frac{dx^{j}}{ds}\frac{dx^{k}}{ds}=f^{i}.
\end{equation}

Eq.(\ref{f1}) yields for null geodesic:
\begin{equation}\label{f7}
 \frac{dy}{ds}=\frac{1}{\Phi}.
\end{equation}
With this condition for metric (\ref{f1}) with 4D line element
(\ref{f2}) fifth force is written as
\begin{equation}\label{f8}
 f^{i}={\mathrm{-}}\frac{g^{ik}}{\Phi}\left(\frac{\partial\Phi}{\partial x^{k}}+
\frac{\partial g_{kj}}{\partial y}u^{j}\right)-\omega u^{i}.
\end{equation}
 The fifth component of Eq. (\ref{f6})
takes the following form:
\begin{equation}\label{f9}
\frac{d^{2}y}{ds^{2}}+\frac{1}{\Phi^{2}}\left(\frac{\partial
g_{ij}}{\partial y}u^{i}u^{j}+2\frac{\partial \Phi}{\partial
x^{i}}u^{i}+\frac{1}{\Phi}\frac{\partial \Phi}{\partial
y}\right)+\omega\frac{dy}{ds}=0
\end{equation}
By substitution (\ref{f7}) in (\ref{f9}) we obtain
\begin{equation}\label{f10}
\omega=-\frac{1}{\Phi}\left(\frac{\partial g_{ij}}{\partial
y}u^{i}u^{j}+\frac{\partial \Phi}{\partial x^{i}}u^{i}\right).
\end{equation}
Then equations (\ref{f8}) are rewritten as
\begin{equation}\label{f11}
 f^{i}={\mathrm{-}}\left(g^{ik}-u^{i}u^{k}\right)\frac{1}{\Phi}\left(\frac{\partial\Phi}{\partial x^{k}}+
\frac{\partial g_{kj}}{\partial y}u^{j}\right).
\end{equation}
Notice that this formula yields
\begin{equation}\label{f12}
 f^{i}u_{i}={\mathrm{-}}\left(\delta^{k}_{l}-u^{k}\right)\frac{1}{\Phi}\left(\frac{\partial\Phi}{\partial x^{k}}+
\frac{\partial g_{kj}}{\partial y}u^{j}\right)=0.
\end{equation}

Let as consider an example, when $\Phi=1$, and metric (\ref{f2})
is orthogonal and conforms to asymmetrically warped space-time:
\begin{equation}\label{13}
ds^{2}= M(y)\tilde{g}_{00}(x^{m})dx^{02}+N(y)\tilde{g}_{ii}(x^{m})
dx^{i2},
\end{equation}
where $M$, $N$ are functions of extra coordinate. The components
of the fifth force (\ref{f11}) are following:
\begin{eqnarray}\label{f14}
f^{0}=\left(\frac{M'}{M}-\frac{N'}{N}\right)(M\tilde{g}_{00}u^{02}-1)u^{0},
\nonumber \\
f^{i}=\left(\frac{M'}{M}-\frac{N'}{N}\right)M\tilde{g}_{00}u^{02}u^{i},
\end{eqnarray}
where $(')$  denotes derivative with respect to $y$, and in second
equation velocities $u^{i}$ conform to the spacelike coordinates.

\begin{itemize}
  \item
This work was supported by MSO.
\end{itemize}

\end{document}